# Lattice reconstruction induced multiple ultra-flat bands in twisted bilayer WSe$_2$

En Li[†], Jin-Xin Hu[†], Xuemeng Feng[†], Zishu Zhou, Liheng An, Kam Tuen Law*, Ning Wang*, Nian Lin*

Department of Physics, The Hong Kong University of Science and Technology, Hong Kong SAR, China

[†]These authors contributed equally to this work

*Correspondence to: phnlin@ust.hk, phwang@ust.hk, phlaw@ust.hk

ABSTRACT:

**Moiré superlattices in van der Waals heterostructures provide a tunable platform to study emergent properties that are absent in the natural crystal form. Twisted bilayer transition metal dichalcogenides (TB-TMDs) can host moiré flat bands over a wide range of twist angles. For twist angle close to 60°, it was predicted that TB-TMDs undergo a lattice reconstruction which causes the formation of ultra-flat bands. Here, by using scanning tunneling microscopy and spectroscopy, we show the emergence of multiple ultra-flat bands in twisted bilayer WSe$_2$ when the twist angle is larger than 57°. The bandwidth, manifested as narrow tunneling conductance peaks, is estimated less than 10meV, which is only a fraction of the estimated on-site Coulomb repulsion energy. The number of these ultra-flat bands and spatial distribution of the wavefunctions match the theoretical predictions incredibly well, strongly evidencing that the observed ultra-flat bands are induced by lattice reconstruction. Our work provides a foundation for further study of the exotic correlated phases in TB-TMDs.**

Single-layer two-dimensional (2D) materials have different physical and chemical properties from bulk materials[1, 2, 3]. When vertically stacking two monolayers of 2D materials with a twist angle (or 2D materials with different lattice constants), moiré superstructures will form and such periodic potential can dramatically modulate the electronic and optical properties[4, 5, 6, 7]. In recent years, an important breakthrough was the experimental observation of correlated insulator and superconductivity in twisted bilayer graphene (TBG)[8, 9]. Near magic twist angles, flat bands emerge near the Fermi level in TBG and promote electron-electron interactions, resulting in strongly correlated phases, such as superconductivity and correlated insulators[8, 9, 10, 11].

Inspired by the discoveries in TBG, flat bands have been predicted in other twisted bilayers with moiré superstructures[12, 13, 14], especially in twisted bilayer transition metal dichalcogenides (TB-TMDs)[14, 15]. Unlike the flat bands in TBG which only emerge at certain twist angles, flat bands with bandwidth about tens of meV appear in a wide range of twisted angles in twisted bilayer TMDs[14, 16]. Since TMDs have large effective masses and relatively strong electron-electron interactions[17, 18], varied flat-band behaviors with twist angle make TB-TMDs a highly tunable platform to study correlated physics[15, 16, 19]. For example, novel topological insulating states have been predicted to emerge in TMD heterobilayer[20] and twisted homobilayers[21]. Evidence of low-energy flat bands and correlating physics, including correlated insulators and generalized Wigner crystal, have been observed in twisted bilayer TMDs by both transport and optical studies[22, 23, 24, 25, 26]. Despite these exciting progresses, there is an urgent need to understand the properties and underlying physical origin of the flat bands.

In this work, we observed multiple ultra-flat bands near valence band (VB) edges in twisted bilayer $WSe_2$ (TB-$WSe_2$) with various twist angles around 60°, by low-temperature scanning tunneling microscopy and spectroscopy (STM/STS) measurements. These ultra-flat bands, with bandwidths estimated to be a few meV only, appear when the twist angle is larger than 57°. The number of the flat bands increases and the bandwidth decreases as the twist angle approaches 60°. Combining with theoretical modeling, we conclude that in addition to the well-known basic moiré bands caused by interlayer hybridization, the ultra-flat bands mainly arise from atomic reconstruction of the moiré superlattices when the twist angle is larger than 57° due to the enhanced interlayer interaction. The ultra-flatness of the bands in the order of meV, makes this system highly susceptible to electron-electron interactions induced effects. Our work lays the foundation for the understanding of the emergence and tunability of ultra-flat bands in bilayer WSe$_2$, which is essential for the understanding of correlated phases in these materials.

**Electronic structure without lattice reconstruction in the sample with 54.1° twist angle:** Different from TBG, twisted bilayer TMDs form two distinct moiré structures for the twist angle ($\theta$) close to 0° and 60° [14, 16], owing to sublattice symmetry breaking in the TMDs. The distinct high symmetry stackings make the electronic structure for twist angle around 0° and 60° significantly different[14, 16], for example, the different spatial distribution of the flat band wavefunctions in 3° and 57.5° TB-WSe$_2$[27]. In this work, we focus on the samples with the twist angle around 60° due to the following: It has been shown by theoretical analysis[28, 29] and transmission electron microscopy (TEM) observations[30, 31] that, at a small misalignment angle, the moiré superlattices undergo a strongly structural reconstruction rather than rigidly rotated crystalline lattices. The reconstruction necessarily modifies the electronic and excitonic properties of the twisted bilayers[32, 33]. Especially, for the twist angle close to 60°, calculations predict that local strains in reconstructed moiré superlattices play an important role in engineering the moiré potential, resulting in multiple energy-separated ultra-flat bands[16]. However, experimental evidence for these ultra-flat bands is still missing. Herein, series of TB-WSe$_2$ with various twist angles around 60° have been fabricated and investigated. The samples cover the transition from un-reconstructed lattice to reconstructed lattice, which allows us to distinguish the physical origin of the flat bands in TB-WSe$_2$ with and without lattice reconstruction.

Fig.1**a** shows an optical microscope image of a 54° TB-WSe$_2$ sample, in which lattice reconstruction does not occur. Two pieces of monolayer WSe$_2$ sit on a highly oriented pyrolytic graphite (HOPG) substrate with a rotation angle of 54°. The stacked region is marked by a red-dashed box, confirming that the stacking region only consists of bilayer WSe$_2$. In the bilayer region, a uniform moiré pattern is resolved by STM, as shown in Fig.S1a. An atomic-resolution STM image shows the topmost Se atoms of WSe$_2$ with a moiré corrugation (Fig.1**b**). Based on the measured moiré period ($\ell$: ~3.2 nm) and atomic lattice (~0.33 nm), the twist angle can be identified to be 54.1°, in good agreement with the alignment angle determined in the sample fabrication. The appearance of the moiré corrugation in STM images originates from various high symmetry stackings. We label the spatial locations as O, A, B (2H stacking), and Br (bridge between A and B). Their corresponding stackings are illustrated at the bottom of Fig.1**b**, according to the structural modeling[14] and previous STM/STS analysis[27]. We carried out the STS measurements to investigate the electronic structure in this moiré superlattice. Fig.S1b is the large-range logarithm of dI/dV spectra acquired at the center of O, A, B, and Br. These spectra consistently show a nearly intrinsic semiconductor bandgap of 2.04 eV. A careful inspection of the site-specific spectra and the spatially-resolved conductance map in Fig.S1c reveals that the VB edge shows moiré dependent features.

We focus on the VB edge. As shown in Fig.1**c**, B, A, and Br sites feature different dI/dV spectra. Site B shows a pronounced peak at −1.14 V (v1) with a full width at half maximum (FWHM) of ~37meV. Site A shows a peak at −1.17 V (v2). As the bridge connecting A and B, Br sites show the residual of the first two peaks and raise the third peak at −1.21 V (v3). To elucidate the spatial distribution of the electronic states with respect to the moiré pattern, we acquired a series of STS spectra across the four high-symmetry sites along the green line in Fig.1**b**, presented as the line-mode conductance map displayed in Fig.1**d**. The v1 (v2) state is mostly distributed in the B (A) region, while the v3 state is located at the Br site. To further resolve the spatial distribution of these electronic states, we performed 2D spectroscopic mapping near the VB edge. Fig.1**e** shows the 2D maps at the selected energies (v1-v3) of the same surface area, showing that the v1 and v2 states are distributed in the B and A regions, respectively. The v3 mainly lies at the Br site, displaying a Kagome-like lattice structure as indicated in Fig.1**e**(v3).

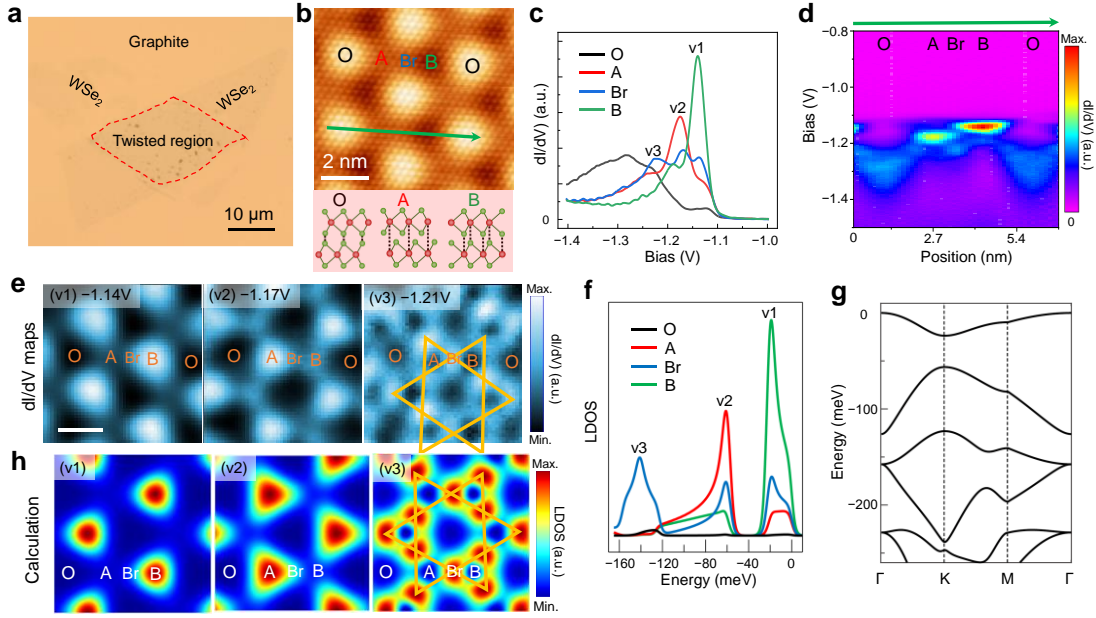

**Figure 1. STM/STS measurement on 54.1° TB-WSe$_2$. a**, Optical image of 54.1° TB-WSe$_2$ sample. The red-dashed box highlights the twisted region. **b**, Top: an atomic-resolution STM topographic image (−1.5 V, 600 pA) of 54.1° TB-WSe$_2$. Bottom: a schematic side view of stacking configurations between two WSe$_2$ layers: O, A, B. Green: Se; Red: W. **c**, dI/dV spectrums (V=−1.5 V, I=500 pA, V$_{mod}$=10mV) near VB edge acquired at the center of four moiré sites. **d**, Conductance map taken along the green line in panel **b**, V=−1.5 V, I=1 nA, V$_{mod}$=10mV. **e**, dI/dV maps at −1.14 V (v1), −1.17 V (v2) and −1.21 V (v3). Scale bar: 2nm. **f**, Calculated LDOS at four moiré sites, showing the site-dependent peak intensity in Moiré lattice. **g**, Calculated electronic band structure plotted in the folded Moiré Brillouin zone. Note that the valence band edge is set at E=0. **h**, Calculated LDOS maps at the energy of v1, v2, and v3, respectively, agree well with experimental results shown in **e**.

To understand the origin of the spectral peaks near the VB edge depicted in Fig.1**b**, we calculated the electronic structure using a continuum model for the Γ valley moiré structure of TB-TMDs (Methods). Fig. 1**f** is the calculated local density of states (LDOS) at four moiré sites, which match very well with the measured site-specific dI/dV spectrums (Fig. 1**c-d**). Fig. 1**g** shows the calculated band structure which reveals a series of moiré bands with bandwidths in the order of tens of meV (v1: ~30 meV, and v2: ~60meV). Furthermore, the spatial distributions of the calculated density of states at the energy of the spectral peaks are displayed as 2D LDOS maps shown in Fig. 1**h**, agreeing very well with measured 2D dI/dV maps (Fig.1**e**). For 54.1° TB-WSe$_2$, the onsite Coulomb repulsion is estimated: $U = e^2/4\pi\epsilon a_m \approx 110$ meV for in-plane dielectric constant $\epsilon = 4$. Such a reasonably large ratio for $U$/bandwidth suggests the possibility of correlated insulator phase at half-filling or fractional-filling of the first moiré band[25, 34].

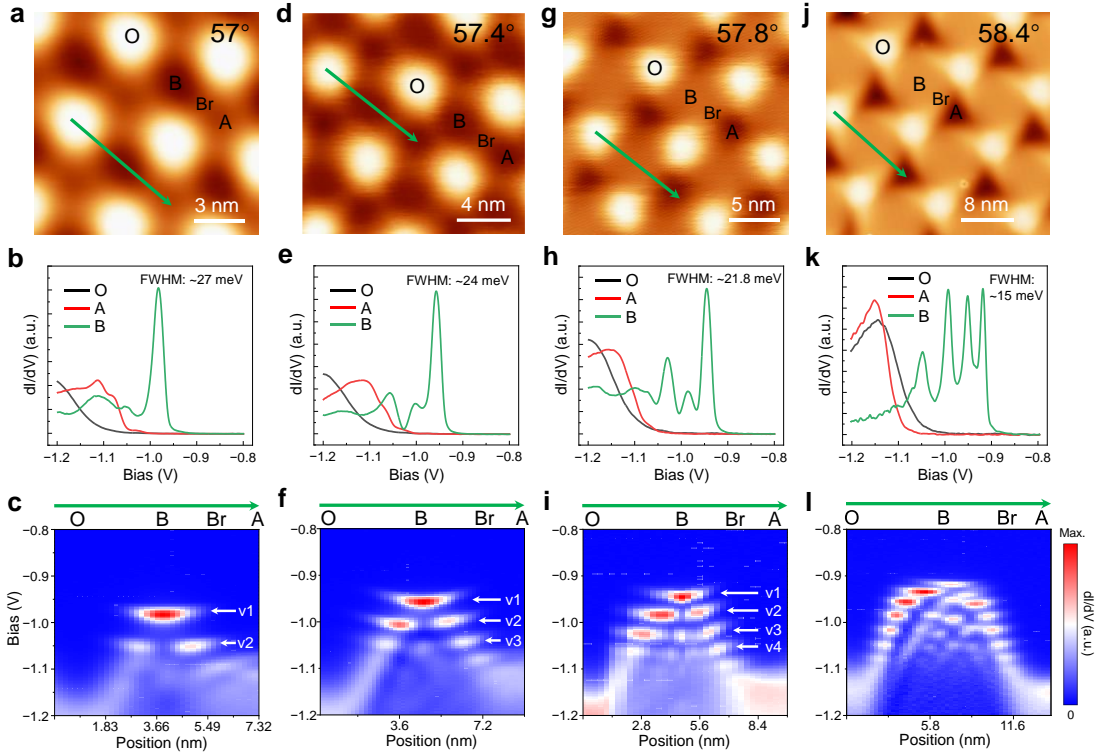

**Figure 2. Spatially dependent spectroscopy of 57°, 57.4°, 57.8°, and 58.4° TB-WSe$_2$. a**, An STM topographic image (−1.4 V, 1.0 nA) of 57° TB-WSe$_2$. **b**, dI/dV spectrums near VB edge acquired at the center of three moiré sites, V=−1.4 V, I=1 nA, V$_{mod}$=10mV. The FWHM is extracted from a Gaussian fit applied to the data. **c**, Conductance map taken along the green line in panel **a**, revealing two B-confined states (marked by white arrows). Similarly, **d**-**f** for 57.4°, V=−1.4 V, I=1 nA, V$_{mod}$=10mV; **g-i** for 57.8°, V=−1.4 V, I=1 nA, V$_{mod}$=10mV; **j-l** for 58.4°, V=−1.3 V, I=1 nA, V$_{mod}$=5 mV.

**Lattice reconstructions at twist angle larger than 57°:** To explore the evolution of moiré-mediated bands against the twist angle, we fabricated and measured the TB-WSe$_2$ samples with twist angles of 57°, 57.4°, 57.8°, and 58.4°. For each moiré pattern, we systematically carried out spatial-dependent STS measurements. Fig.2**b**, **e**, **h**, and **k** display the dI/dV spectrums near the VB edge acquired at the center of three moiré sites, revealing the isolated sharp band-edge peaks (in green) at B sites near the VB edge. Fig.2**c**, **f**, **i**, and **l** show the line-mode conductance maps acquired point-by-point along the green line in the corresponding STM image in Fig.2, displaying the evolution of the electronic states near the VB edge at gradually increased twisted angles. The 57° TB-WSe$_2$ ($\ell$: ~6.3 nm) features two states (v1 and v2) near the VB edge that are separated by ~66 meV in energy, as marked by two white arrows in Fig.2**c**. Both states are localized in region B but their spatial distributions are very different: v1 is distributed at the center region while v2 is distributed at the two sides having a central "node". As the twist angle further increased to 57.4° ($\ell$: ~7.2 nm) and 57.8° ($\ell$: ~8.6 nm), we observed three (v1, v2, and v3) and four (v1, v2, v3, and v4) energy-separated states that were localized in the region B, respectively. The distributions of v1 and v2 are similar to those of the 57.0° sample. The v3 state has two nodes and the v4 state has three nodes. As will be shown below, these nodes are related to the quantum well states caused by lattice reconstruction. For the 58.4° TB-WSe$_2$ sample ($\ell$: ~11.8 nm), a sequence of states emerges in region B as shown in Fig.2**k**. Apart from the increased states numbers, the energy separation of observed states decreases as the twist angle approaches 60°. Note that the slight concave-up streaks of these states distribution are likely due to tip-induced band bending (TIBB) effects, similar behavior has been observed in aligned MoS$_2$/WSe$_2$ heterobilayer[35].

The continuum model with the $V_M(r)$ moiré potential (Methods) can well describe the states in 54.1° TB-WSe$_2$, however, we find it fails to interpret the quantum-well-like states observed in the 57°-58.4° samples. This indicates that the multiple ultra-flat bands in the 57°-58.4° samples have distinct origins from the 54.1° sample. As revealed by early works[28, 30], for $\theta > 57°$, TB-TMDs feature lattice reconstruction and domain formation. Calculations demonstrate that local strains caused by the reconstruction result in the formation of the triangular potential wells in moiré structure, which will give a series of localized quantum-well states[16]. Therefore, we adopt a reconstructed confining potential ($V_C$) to describe the atomic reconstruction and rewrite the continuum model, as described in Methods and Supplementary Section 4. Fig.3 shows the calculated band structure (left) and the LDOS distribution across the O-B-A-O sites (right), for 57° (**a**), 57.4° (**b**), 57.8° (**c**), and 58.4° (**d**) TB-WSe$_2$. Multiple energy-separated flat bands emerge at the VB edge and are nearly dispersionless with bandwidth less than 1 meV. These ultra-flat bands are confined in region B, showing the characteristics of the quantum-well-like states. The numbers and energy separation of ultra-flat bands evolve with the twist angles. The relative energy levels of the experimentally resolved quantum-well-like states are plotted in orange lines. One can see that the energy levels, as well as the spatial distributions of the quantum-well-like states, agree well with the experimental data shown in Fig.2**c**, **f**, **i**, and **l**.

**Ultra-flat bands induced by lattice reconstruction:** Here we discuss the bandwidth of the flat bands. The calculated flat bands for twist angles greater than 57° are nearly dispersionless, while 54.1° TB-WSe$_2$ hosts a flat band with a bandwidth of ~30 meV (v1). In our STS measurement, the energy resolution is given by $\Delta E = \sqrt{(3.5kT)^2 + (2.5V_{mod})^2}$[35]. Since we use a modulation voltage of V$_{mod}$=10 mV for 54.1°-57.8° and 5 mV for 58.4° to distinguish peaks, the corresponding energy resolution at 5.3 K is 25 meV and 12.6 meV, respectively. For the 54.1° sample, a slightly larger peak width (FWHM: ~37 meV) could be attributed to the small dispersion of the first band as shown in Fig.1**g**. In contrast, for 57°-58.4° TB-WSe$_2$, many of B-confined peaks are seen to have width very close to the value of STS resolution. As shown in Fig.2**b**, **e**, **h**, and **k**, FWHM of the first band-edge peak in center of B region is ~27-21.8 meV for 57°-57.8° samples (V$_{mod}$= 10 mV) and ~15 meV for 58.4° (V$_{mod}$= 5 mV). Hence, the observed widths of these peaks are mainly produced by the modulation. By fitting the obtained tunneling spectra, as described in Supplementary Section 2, we estimate the intrinsic bandwidth of the ultra-flat bands in the 57°, 57.4°, 57.8° and 58.4° twisted sample to be 10.2, 8.8, 7.5, and 5.8 meV, respectively. Such narrow bandwidth of these ultra-flat bands, which is only a fraction of the estimated on-site Coulomb repulsion energy, makes this system highly susceptible to electron-electron interaction induced effects.

Furthermore, we perform dI/dV mapping to study the spatial distribution of the observed sharp peaks in Fig.2. As shown in Fig.3**c**, **f**, **i**, **l**, dI/dV maps of the states are sequentially consistent for 57°-58.4° TB-WSe$_2$. We note that the wavefunctions of the flat bands at different energy have very different spatial distributions, associated with the observed nodes in line-mode conductance maps in Fig.2. We have calculated the wavefunction patterns of the ultra-flat bands with lattice reconstruction and the results are depicted in Fig.3**b**, **e**, **h**, and **k**, showing that v1 appears as a dot at the center of region B, v2 appears as a three-dot triangle, v3 and v4 comprise more dots following the triangular symmetry. The calculated spatial distribution of the wavefunctions matches the dI/dV maps incredibly well. This provides direct and convincing experimental evidence that the ultra-flat bands are originated from the lattice reconstruction.

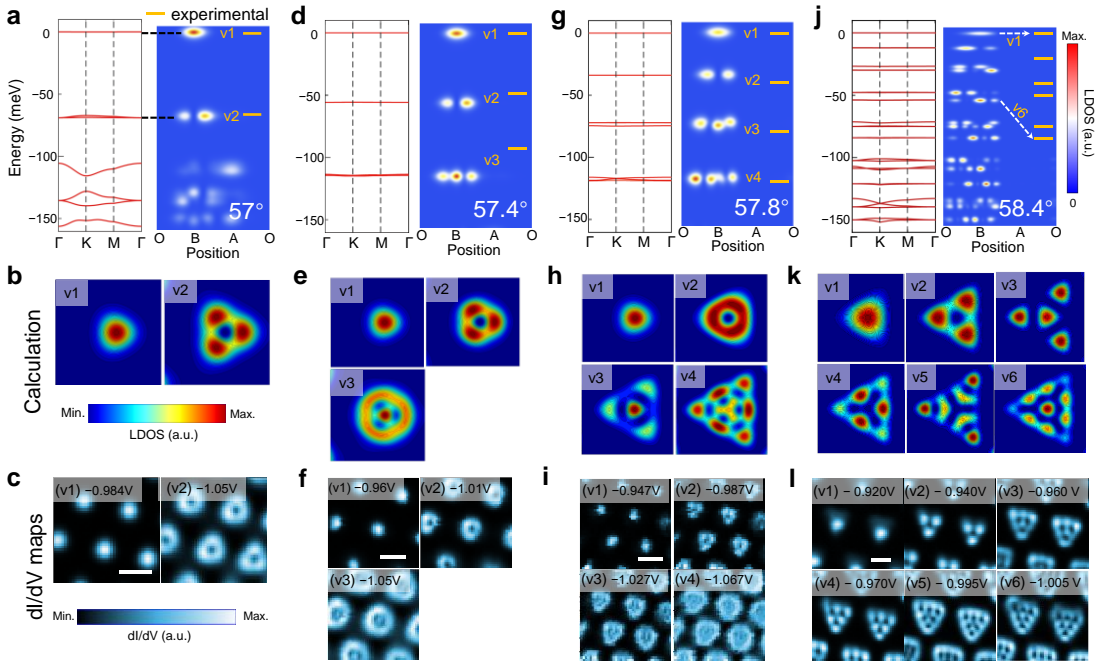

**Figure 3. Evolution of calculated flat bands with twist angle. a**, Calculated band structure (left) and the LDOS across the O-B-A-O sites (right) for 57° TB-WSe$_2$, showing the ultra-flat bands and corresponding spatial distributions. The orange lines mark the relative energy levels of the experimental states. Note that the valence band edge is set at E=0. **b**, Calculated

wavefunction patterns of the flat bands labeled in **a**. **c**, Measured dI/dV maps for 57° TB-WSe$_2$ at the energies of B-confined states in Fig. 2**c**. Scale bar: 4 nm. Similarly, **d**-**f** for 57.4°, scale bar in **f**: 4 nm; **g**-**i** for 57.8°, scale bar in **i**: 6 nm; and **j**-**l** for 58.4°, scale bar in **l**: 5 nm. **k** and **l** display the calculated and experimental LDOS maps of first six states in 58.4° TB-WSe$_2$, respectively.

**Lattice reconstruction induced evolution of the morphology, moiré potential, and flat bands:** The exceptionally high degree of agreement between the STS measurements and the modeling suggests that the flat bands in the TB-WSe$_2$ can be attributed to the emergence of a moiré potential associated with lattice reconstruction. For the 54.1° sample, its potential term $V_M(r)$ is from inhomogeneity of the interlayer hybridization in the moiré superlattice. As the twist angle further approaches 60°, lattice reconstruction happens when energy gain from the formation of favorable stacking overcomes elastic energy cost of strain produced by the local in-plane displacements[29]. Different from TBG and TB-TMDs around 0°, there is only one low-energy stacking (2H) in TB-TMDs with $\theta$ close to 60°[28, 36]. In real space, the 2H stacking domains (B regions) expand overwhelmingly as twist angle approaches 60° and transform from an equilateral triangle to a Reuleaux triangle[16, 28], as illustrated in Fig.4**a**. In our STM images shown in Fig.4**b**, the reduced size of O regions and increased B regions corroborates this phenomenon. This structural transformation of the moiré superlattice strongly influences the electronic structure, specifically, forming the multiple ultra-flat bands. Local strain caused by in-plane relaxations, results in the formation of modulating confining potential[16]. As displayed in Fig.4**c-d**, without reconstruction (54.1°), the moiré potential profile modulates gradually and reaches maximum at the B site, which produces v1 state. The local maximum at the A site produces the v2 state. In sharp contrast, with the lattice reconstruction (58.4°), the potential profile features periodic quantum wells that give a series of confining states (v1, v2, v3, … etc.) in the B regions depending on the size and potential depth of the well.

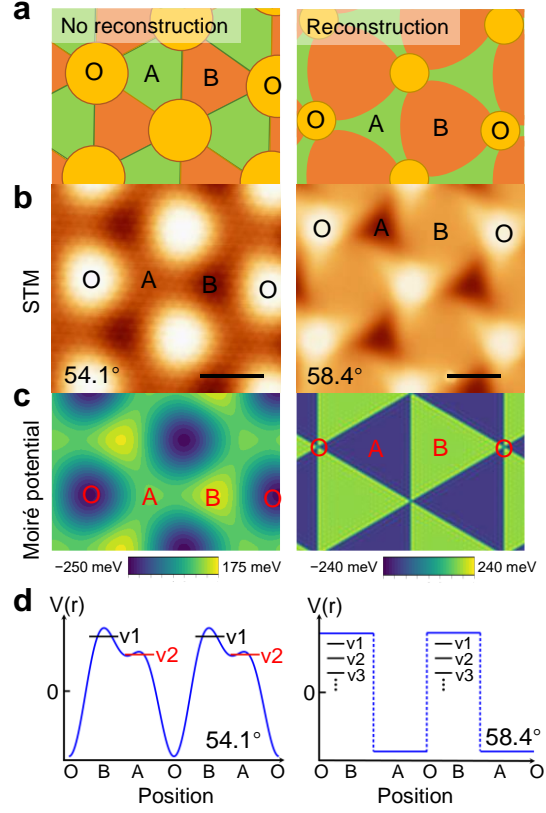

**Figure 4. Moiré potential evolution due to lattice reconstruction. a**, Illustration of the moiré pattern with and without reconstruction (right and left, respectively). **b**, STM image of 54.1° (left, V=−1.4 V, I=1 nA) and 58.4° (right, V=−1.25 V, I=1 nA) TB-WSe$_2$, showing the topographic difference due to lattice reconstruction. The scale bar is 2 nm (54.1°) and 6 nm (58.4°), respectively. **c**, Moiré potential for 54.1° (left) and 58.4° (right) TB-WSe$_2$. **d,** Illustration of moiré potential profile for 54.1° (left) and 58.4° (right) TB-WSe$_2$ and the corresponding confined flat bands.

The evolution of the moiré potential $V_M$ and confining potential $V_C$ against the twisted angles used in our model calculations is shown in Fig.S4. The gradual variation of the two potentials indicates a transition from almost rigid bilayers, where the interlayer hybridization dominates, to lattice reconstruction with the relaxation-induced confining potential dominates. We notice there is a peak around −1.1V at the center of A regions in 57° TB-WSe$_2$ (Fig.2**b**), besides the two B-confined states. This indicates the cooperative interactions between moiré potential $V_M$ and confining potential $V_C$. Moreover, we find that this transition accompanies the enhanced interlayer interaction in the bilayer WSe$_2$. As shown in Fig. S2, the states away from VB/CB edges vary with the moiré structure. The peaks at ~ −1.8V and 1.3-2.0V show an energy downward shift at A sites. This downward shift becomes more pronounced as the twist angle changes from 54.1° to 58.4°, evidencing stronger interlayer interaction as the twisted angle increases.

In summary, this study demonstrates two distinct mechanisms for the emergence of the flat bands in TB-WSe$_2$ with twisted angles around 60°: At smaller twisted angles (for instance 54.1°), the interlayer hybridization leads to two spatially separated flat bands with bandwidths in the order of tens of meV. At larger twisted angles ($\theta >57°$), lattice reconstruction accompanied with strong interlayer interaction creates the triangular quantum wells that host multiple energy-separated ultra-flat bands with bandwidths of only a few meV. The observed ultra-flat bands with tunable numbers and energy separation provide practical guidance for further realization of correlated states when the Fermi energy is properly tuned to the ultra-flat bands. We expect that the large on-site Coulomb repulsion energy to bandwidth ratio will result in interesting correlated phases in these lattice reconstructed systems.

## Methods

### Sample fabrication.

Our 54°~58.4° TB-WSe$_2$ samples were fabricated by the "tear and stack" method[37], based on the PDMS stack and the precisely rotational control stage (Mechanical accuracy is ±0.5°). The WSe$_2$ bulks are bought from HQ Graphene. The monolayer WSe$_2$ flakes with large size are exfoliated onto the SiO$_2$/Si wafer, which always have natural cracks, convenient for the same flake "tear and stack" process. Use the polydimethylsiloxane (PDMS) stack covered by the poly(bisphenol A carbonate) (PC) thin film to tear half (or part) of the monolayer flake, and stack onto the remnant part monolayer flake with a controlled interlayer twist angle near 60°. Then, the TB-WSe$_2$ structures are transferred onto the freshly cleaved graphite substrate.

### STM measurements.

STM/STS experiments were carried out in a commercial ultra-high vacuum low-temperature STM system (CreaTec) with a base pressure of 1.0×10$^{-10}$ mBar. All the STM/STS measurements were performed at 5.3 K using a chemically etched tungsten tip. Before STM measurement, the samples were annealed at ~ 250 °C for over 3 hours to remove possible adsorbates. STM images were acquired in the constant-current mode and the bias voltages refer to the sample with respect to the STM tip. The dI/dV spectra were collected by using the standard lock-in technique with a voltage modulation of 5-10 mV and frequency of 787.3 Hz.

### Theoretical Modeling of Moiré band structure

We construct the continuum model for AB stacking twisted WSe$_2$ bilayers. For AB stacked bilayer WSe$_2$, the valence band maxima is always at Γ point not K point. Neglecting spin-orbit coupling, which vanishes at the Γ-point by Kramer's theorem, we obtain the following simple single-band $k \cdot p$ Hamiltonian:

$$H = \frac{\hbar^2 k^2}{2m^*} + V_M(\boldsymbol{r})$$

$$V_M(\boldsymbol{r}) = 2V_0 \sum_i \cos(\boldsymbol{g}_i \cdot \boldsymbol{r} + \phi)$$

$m^* \approx 1.2 m_e$ is the effective mass at $\Gamma$ pocket and $V_0, \phi$ are parameters which describe the moiré potential term. And $\boldsymbol{g_i}$ are moiré reciprocal vectors. We choose a typical set of parameters $(V_0, \phi) = (40 meV, 170°)$ for $\theta = 54.1°$, which reproduces the experimental LDOS structure both qualitatively and quantitatively.

When the twist angle is close to 60°, the WSe$_2$ bilayers undergo an atomic reconstruction and the moiré potential should be changed due to the existence of triangular networks[16].

We use the reconstructed confining potential:

$$V_C(\boldsymbol{r}) = -2V_1 \sum_{n=1}^{\infty} \sum_i \frac{1}{n} \sin(n\boldsymbol{g_i} \cdot \boldsymbol{r})$$

We choose $V_1 = 50$ meV for $\theta = 58.4°$.

The formula of local density of states:

$$\psi_{n\boldsymbol{k}}(\boldsymbol{r}) = \sum_{\boldsymbol{G}} u_{n\boldsymbol{k}}(\boldsymbol{G}) e^{i(\boldsymbol{k}+\boldsymbol{G})\cdot \boldsymbol{r}}$$

$$D(\boldsymbol{r}, E) = \sum_{n,\boldsymbol{k}} |\psi_{n\boldsymbol{k}}(\boldsymbol{r})|^2 \delta(E - E_{n\boldsymbol{k}})$$

## Acknowledgements

This work is supported financially by the Hong Kong UGC (C6012-17E), the National Key R&D Program of China (2020YFA 0309600/0309602) and the Research Grants Council (RGC) of Hong Kong (Project No. 16300720).

## Author contributions

N.L., N.W. and K.-T.L. supervised the work. E.L. performed the STM/STS experiments. X. F and L. A. fabricated the samples with the assistance of Z. Z. J.-X.H. computed the electronic structure under the supervision of K.-T.L. E.L., and J.-X.H. analyzed the the experimental data together with computational results. E.L., J.-X.H., N.L., N.W. and K.-T.L. wrote the manuscript with input from all other authors.

We have no competing Interests.